\documentstyle[preprint,prd,aps,psfig]{revtex}
\tightenlines
\newcommand{\be}{\begin{eqnarray}}
\newcommand{\ee}{\end{eqnarray}}
\newcommand{\bea}{\begin{eqnarray}}
\newcommand{\eea}{\end{eqnarray}}
\newcommand{\ba}{\begin{array}}
\newcommand{\ea}{\end{array}}

\newcommand{\gam}{{\mbox{$\vec{\gamma}$}}}

\newcommand{\Op}{{\cal O}}

\newcommand{\lw}[1]{\smash{\lower1.6ex\hbox{#1}}}
\newcommand{\nabd}{\loarrow{D}}
\newcommand{\nabr}{\rlap{\hbox{$D$}}
                   \raise 8 pt \hbox{$\hspace{-0.05cm}\leftarrow$}}
\def\simgt{\rlap{\lower 3.5 pt\hbox{$\mathchar \sim$}}
           \raise 1pt \hbox {$>$}}
\def\simlt{\rlap{\lower 3.5 pt\hbox{$\mathchar \sim$}}
           \raise 1pt \hbox {$<$}}
\begin{document}
\draft
\preprint{\vbox{\hbox{HUPD-9903}}}
\title{
  A lattice NRQCD calculation of the $B^0$-$\bar{B}^0$ mixing
  parameter $B_B$
  }
\author{
  S. Hashimoto$^{\mbox{\scriptsize\ a}}$,
  K-I. Ishikawa$^{\mbox{\scriptsize\ b}}$,
  H. Matsufuru$^{\mbox{\scriptsize\ b}}$,
  T. Onogi$^{\mbox{\scriptsize\ b}}$,
  and
  N. Yamada$^{\mbox{\scriptsize\ b}}$
}
\address{
  $^{\mbox{\scriptsize\ a}}$
  High Energy Accelerator Research Organization (KEK),
  Tsukuba 305-0801, Japan\\
  $^{\mbox{\scriptsize\ b}}$
  Department of Physics, Hiroshima University,
  Higashi-Hiroshima 739-8526, Japan
}
\maketitle
\begin{abstract}
We present a lattice calculation of the $B$ meson
$B$-parameter $B_B$ using the NRQCD action.
The heavy quark mass dependence is explicitly studied over a
mass range between $m_b$ and $4m_b$ with the $O(1/m_Q)$ and
$O(1/m_Q^2)$ actions.
We find that the ratios of lattice matrix elements
$\langle O_N^{lat} \rangle/\langle A_0^{lat} \rangle^2$ and
$\langle O_S^{lat} \rangle/\langle A_0^{lat} \rangle^2$,
which contribute to $B_B$ through mixing, have
significant $1/m_Q$ dependence while that of the leading operator
$\langle O_L^{lat} \rangle/\langle A_0^{lat} \rangle^2$
has little $1/m_Q$ effect.
The combined result for $B_B(m_b)$ has small but non-zero
mass dependence, and the $B_B(m_b)$ becomes smaller by 10\%
with the $1/m_Q$ correction compared to the static result.
Our result in the quenched approximation at $\beta$=5.9 is
$B_{B_d}$(5 GeV) = 0.75(3)(12), where the first error is
statistical and the second is a systematic uncertainty.
\end{abstract}
\pacs{PACS number(s): 12.38.Gc, 12.39.Hg, 13.20.He, 14.40.Nd}
\section{Introduction}
\label{sec:intro}

The constraints on the unitarity triangle of the
Cabibbo-Kobayashi-Maskawa~(CKM) matrix can provide us with one of the
most crucial information on the physics beyond the Standard
Model~\cite{CKM_unitarity}.
However, due to large theoretical or experimental uncertainties,
the current bound is too loose to test the Standard Model
or new physics.
The $B^0$-$\overline{B^0}$ mixing sets a constraint on
$|V_{td}V_{tb}^*|$ through the currently
available experimental data on the mass difference between
two neutral $B$ mesons $\Delta M_{B}$= 0.477$\pm$0.017
ps$^{-1}$ \cite{LEPBOSC}.
The experimental achievement is impressive as the error is
already quite small $\sim$ 4\%.
Theoretical calculation to relate $\Delta M_{B}$ to
$|V_{td}V_{tb}^*|$, on the other hand, involves a large 
uncertainty in the $B$ meson matrix element
$\langle\overline{B^0}|{\cal O}_{L}|B^0\rangle$, which requires 
a method to calculate the non-perturbative QCD effects.

Lattice QCD is an ideal tool to compute such
non-perturbative quantities from first principles.
There has been a number of calculations of the $B$ meson
decay constant $f_{B}$ and the $B$-parameter $B_{B}$
($B_{B}$ describes the matrix element through
$\langle\overline{B^0}|{\cal O}_{L}|B^0\rangle$ = 
$(8/3) B_{B} f_{B}^2M_{B}^2$).
The calculation of $f_{B}$ is already {\it matured}
at least in the quenched approximation~\cite{draper}. 
Major systematic errors are removed by introducing the
non-relativistic effective actions and by improving the
action and currents.
Remaining $a$ (lattice spacing) dependent systematic error
is confirmed to be small, and in some papers the continuum
extrapolation is made.
A recent review~\cite{draper} summarized the value of $f_{B}$
as $f_{B}$= 165$\pm$20 MeV within the quenched approximation.

An essential ingredient of these calculations is the use of
the non-relativistic effective actions.
Since the $b$-quark mass in lattice unit $a m_b$ is large
for the typical lattices used for simulations,
the relativistic (Wilson-type) actions could suffer from 
a large discretization error of order $am_b$ or $(am_b)^2$.
The non-relativistic QCD (NRQCD)~\cite{NRQCD}, on the other
hand, is formulated as an expansion in $\mbox{\boldmath$p$}/m_Q$.
In the heavy-light meson system, where the typical spatial 
momentum scale is $\Lambda_{QCD}$, the error from the 
truncation of higher order terms is controllable.
The calculation of $f_B$ is now available to order
$1/m_b^2$, and it is known that the contribution of
$O(1/m_b)$ is significant ($\sim -20\%$) while that of
$O(1/m_b^2)$ is small ($\sim -3\%$)~\cite{Our}.
In addition, the calculations based on the Fermilab approach
for heavy quark~\cite{EKM,sofarfb} agree with the NRQCD
results~\cite{jlqcd_nrfb} including their $1/m_Q$
dependence. 
These results make us confident about the non-relativistic
effective action approaches in Lattice QCD.
Now that the computation of $f_{B}$ is established,
the next goal is to apply the similar technique
to the computation of $B_{B}$.

The lattice calculations of the $B$-parameter have been done in the
infinitely heavy quark mass (static) limit~\cite{UK,GM,CDM},
and the results are in reasonable agreement with each other.
There is, however, some indication that the $1/m_b$
correction would be non-negligible from the study with
relativistic quark actions~\cite{BBM,LL}.
Their results show that there is small but non-zero negative
slope in the $1/m_Q$ dependence of $B$-parameter, but it is not
conclusive because of the possible systematic uncertainty
associated with the relativistic quark action for heavy quark.
The purpose of this work is to study the $1/m_Q$ dependence
of $B_{B}$ by explicitly calculating it with the NRQCD
action at several values of $1/m_Q$.
Our result confirms the previous works~\cite{BBM,LL}:
there is a small negative slope in $B_{B}$.
In addition, we find that the slope comes from the large
$1/m_Q$ dependence of $B_N^{lat}$ and $B_S^{lat}$, which are matrix
elements of non-leading operators.
For the observed $1/m_Q$ dependence of the lattice matrix elements,
qualitative explanations are given in Discussion section
using the vacuum saturation approximation.

The perturbative matching of the continuum and lattice
operators introduces a complication to the analysis.
Since the one-loop coefficients for four-quark operators are
not known yet for the NRQCD action, we use the
coefficients in the static limit in Refs.~\cite{FHH,BP,PS,GR,IOY}.
The systematic error associated with this approximation is
of $\alpha_s/(am_Q)$ and expected to be small compared to the $1/m_Q$
correction itself.
This and other systematic errors are discussed in detail in
the Discussion section.

This paper is organized as follows.
In the next section, we summarize our matching procedure.
The simulation method is described in section~\ref{sec:method},
and our analysis and results are presented
in section~\ref{sec:results}.
The results are compared with the vacuum saturation
approximation on the lattice in section~\ref{sec:VSA}, and we estimate
the remaining uncertainties in section~\ref{sec:uncertainty}.
Finally our conclusion is given in section~\ref{sec:conclusion}.
An early result of this work was presented in Ref.~\cite{YA}.

\section{Perturbative matching}
\label{sec:pertmat}

In this section, we give our notations and describe the
perturbative matching procedure.
The mass difference between two neutral $B_q$ mesons is given by
\begin{equation}
  \label{dm}
  \Delta M_{B_q} =
  |V_{tb}^*V_{tq}|^2
  \frac{G_F^2 m_W^2}{16 \pi^2 M_{B_q}}
  S_0(x_t) \eta_{2B} 
  \left[ \alpha_s(\mu) \right]^{-6/23}
  \left[ 1 + \frac{\alpha_s(\mu)}{4\pi}J_5 \right]
    \langle \overline{B_q^0}|{\cal O}_L(\mu)
    | B_q^0\rangle,
\end{equation}
where $q=d$ or $s$ and $S_0(x_t)$~($x_t=m_t^2/m_W^2$) and $\eta_{2B}$
are so-called Inami-Lim function and the short distance QCD
correction, respectively.
Their explicit forms can be found in Ref.~\cite{BBL}.
${\cal O}_L(\mu)$ is a $\Delta B$=2 operator
\begin{equation}
  \label{eq:O_L}
  {\cal O}_L(\mu)
  =\bar{b}\gamma_{\mu}(1-\gamma_5)q
   \bar{b}\gamma_{\mu}(1-\gamma_5)q,
\end{equation}
renormalized in the $\overline{\rm MS}$ scheme with the naive
dimensional regularization~(NDR). 
$J_{n_f}$ is related to the anomalous dimension at the
next-to-leading order with $n_f$ active flavors as 
\begin{equation}
  \label{eq:J_5}
  J_{n_f} = \frac{\gamma^{(0)}\beta_1}{2\beta_0^2}
          - \frac{\gamma^{(1)}}{2\beta_0},
\end{equation}
where
\begin{equation}
  \label{eq:anomalous_dimensions}
  \begin{array}{cc}
    \displaystyle \beta_0 = 11 - \frac{2}{3} n_f, \quad&
    \displaystyle \beta_1 = 102 - \frac{38}{3} n_f, \\
    \displaystyle \gamma^{(0)} = 4, &
    \displaystyle \gamma^{(1)} = -7 + \frac{4}{9} n_f.
  \end{array}
\end{equation}
$n_f$=5 when $\mu$ is greater than or equal to the $b$ quark
mass. 

The $B$-parameter $B_{B_q}$ is defined through
\begin{equation}
  \label{eq:B_B_definition}
  B_{B_q}(\mu) =
  \frac{\langle\overline{B_q^0}|{\cal O}_L(\mu)|B_q^0\rangle}{
    \frac{8}{3} 
    \langle\overline{B_q^0}|A_{\mu}|0\rangle
    \langle 0|A_{\mu}|B_q^0\rangle },
\end{equation}
where $A_{\mu}$ denotes the axial-vector current
$\bar{b}\gamma_{\mu}\gamma_5 q$.
The renormalization invariant $B$-parameter is defined
by
\begin{equation}
  \hat{B}_{B_q}
  = \left[\alpha_s(\mu)\right]^{-\frac{6}{23}}
    \left[ 1+\frac{\alpha_s(\mu)}{4\pi}J_5 \right]
    B_{B_q}(\mu),
\end{equation}
which does not depend on the arbitrary scale $\mu$ up to the
next-to-leading order. 
The scale $\mu$ is conventionally set at the scale of
$b$-quark mass $\mu=m_b$.

In order to calculate the matrix element 
$\langle\overline{B_q^0}|{\cal O}_L(m_b)|B_q^0\rangle$
on the lattice, we have to connect the operator 
${\cal O}_L(m_b)$ defined in the continuum renormalization
scheme with its lattice counterpart.
The matching coefficients can be obtained by requiring
the perturbative quark scattering amplitudes at certain momentum
with continuum $O_L$ operator and with lattice four fermi operators 
should give identical results. 

At the one-loop level the matching gives the following relation
\begin{eqnarray}
    \label{eq:matching_3}
    {\cal O}_L(m_b) & = &
    \left( 1 + \frac{\alpha_s}{4\pi}[4\ln(a^2 m_b^2) + D_L - 14 ] 
    \right) {\cal O}_L^{lat}(1/a) \nonumber \\
    & &
    + \frac{\alpha_s}{4\pi} D_R {\cal O}_R^{lat}(1/a)
    + \frac{\alpha_s}{4\pi} D_N {\cal O}_N^{lat}(1/a)
    + \frac{\alpha_s}{4\pi} D_S {\cal O}_S^{lat}(1/a),
    \nonumber 
\end{eqnarray}
  where ${\cal O}_{\{L,R,N,S\}}^{lat}$ denotes the naive local
  operators defined on the lattice in which the light quarks are not
  rotated. Their explicit forms are the following,
\begin{eqnarray}
    \label{eq:O_R_and_O_N}
    {\cal O}_R & = &
      \bar{b}\gamma_{\mu}(1+\gamma_5)q
      \bar{b}\gamma_{\mu}(1+\gamma_5)q, \nonumber \\
    {\cal O}_N & = &
      \bar{b}\gamma_{\mu}(1-\gamma_5)q
      \bar{b}\gamma_{\mu}(1+\gamma_5)q
    + \bar{b}\gamma_{\mu}(1+\gamma_5)q
      \bar{b}\gamma_{\mu}(1-\gamma_5)q
      \nonumber \\ & &
    + 2 \bar{b}(1-\gamma_5)q \bar{b}(1+\gamma_5)q
    + 2 \bar{b}(1+\gamma_5)q \bar{b}(1-\gamma_5)q, \nonumber \\
    {\cal O}_S & = &
      \bar{b}(1-\gamma_5)q \bar{b}(1-\gamma_5)q.
\end{eqnarray}
Unfortunately, the one-loop coefficients $D_{\{L,R,N\}}$ 
for the NRQCD heavy and  $O(a)$-improved light quark action~\cite{SW}
are not known yet. In this work, we use the
one-loop coefficients in the static limit~\cite{BP,PS,GR,IOY},
\begin{equation}
    \label{eq:D_i}
    D_L = -21.16, \qquad D_R = -0.52, \qquad D_N = -6.16,
    \qquad D_S = -8.
\end{equation}
The systematic error associated with this approximation
is at most $\alpha_s/(am_Q)$, since the NRQCD action's
$m\rightarrow\infty$ limit agrees with the static
action.
The numerical size of the error is discussed later.

The matching of the axial-vector current appearing in the
denominator of Eq.~(\ref{eq:B_B_definition}) can be done in a
similar manner~\cite{BP,GH,Duncan}
\begin{eqnarray}
  \label{eq:matching_A4_1}
  A_0 & = & 
  \left( 1 + \frac{\alpha_s}{4\pi} [2\ln(a^2m_b^2)+D_A-\frac{8}{3}]
	\right) A_0^{lat}(1/a),
\end{eqnarray}
where $A_0^{lat}$ is defined on the lattice, and the matching
coefficient $D_A$ in the static limit $D_A=-13.89$.

In calculating the ratio of Eq.~(\ref{eq:B_B_definition}) a large
cancellation of perturbative matching corrections takes place
between the numerator and denominator, since the large wave
function renormalization coming from the tadpole
contribution in the lattice theory is the same.
To make this cancellation explicit we consider the matching
of a ratio 
$B_B(m_b)=\langle{\cal O}_L(m_b)\rangle
/(8/3)\langle A_0 \rangle^2$ itself,
\begin{eqnarray}
  \label{eq:matching_ratio}
  B_B(m_b) & = &
    Z_{L/A^2}(m_b;1/a) B_L^{lat}(1/a)
  + Z_{R/A^2}(m_b;1/a) B_R^{lat}(1/a) \nonumber \\
  & &
  + Z_{N/A^2}(m_b;1/a) B_N^{lat}(1/a)
  + Z_{S/A^2}(m_b;1/a) B_S^{lat}(1/a),
\end{eqnarray}
where 
$B_{\{L,R,N,S\}}^{lat}(1/a)=
\langle{\cal O}_{\{L,R,N,S\}}^{lat}(1/a)\rangle
/(8/3)\langle A_0^{lat}(1/a) \rangle^2$
and the $B$ and $\overline{B}$ states are understood for the
expectation values $\langle\cdots\rangle$ as in
Eq.~(\ref{eq:B_B_definition}). 
Then the coefficients become
\begin{eqnarray}
  \label{eq:matching_ratio_coeff_L}
   Z_{L/A^2}(m_b;1/a) & = & 
     \left( 1 + \frac{\alpha_s}{4\pi} (D_L-2D_A-\frac{26}{3}) \right), \\
  \label{eq:matching_ratio_coeff_R}
   Z_{R/A^2}(m_b;1/a) & = & 
      \frac{\alpha_s}{4\pi} D_R, \\
  \label{eq:matching_ratio_coeff_N}
   Z_{N/A^2}(m_b;1/a) & = & 
     \frac{\alpha_s}{4\pi} D_N, \\
  \label{eq:matching_ratio_coeff_S}
   Z_{S/A^2}(m_b;1/a) & = & \frac{\alpha_s}{4\pi} D_S.
\end{eqnarray}

The
Eqs.~(\ref{eq:matching_ratio_coeff_L})-(\ref{eq:matching_ratio_coeff_S})
are used in the following analysis to obtain $B_B(m_b)$.
Numerical values of $Z_{\{L,R,N,S\}/A^2}(m_b;1/a)$ are given
in Table~\ref{tab:matchfac} for the lattice parameters in our
simulation.
For the coupling constant $\alpha_s$ in
Eqs.~(\ref{eq:matching_ratio_coeff_L})-(\ref{eq:matching_ratio_coeff_S})
we use the V-scheme coupling~\cite{LM} with $q^*=\pi/a$ or
$q^*=1/a$.
At $\beta$=5.9 those are $\alpha_V(\pi/a)$=0.164 and
$\alpha_V(1/a)$=0.270.
The tadpole improvement \cite{LM} does not make any effect
on the ratio of Eq.~(\ref{eq:matching_ratio}), since the tadpole
contribution cancels between the numerator and denominator.
The $b$-quark mass scale $m_b$ is set to 5 GeV as usual.

In the previous works in the static approximation~\cite{UK,GM,CDM}, 
the leading and the next-to-leading logarithmic corrections are
resummed to achieve better control in the running from $m_b$ to $1/a$.
In this paper we use the one-loop formula without the resummation for
simplicity.
This does not introduce significant error, since the mass scale
difference between $m_b$ and $1/a$ is small and the effect of the
resummation is not important.
In Appendix~\ref{sec:appendix} we compare the $Z$ factors with and
without the resummation.

We determine the heavy-light pseudo-scalar meson mass $M_P$
from the binding energy $E^{\rm bin}$ measured in the simulation
using a formula 
\begin{equation}
  \label{eq:mass_renorm_1}
  a M_P = Z_m am_Q - E_0 + a E^{\rm bin},
\end{equation}
where $Z_m$ and $E_0$ are the renormalization constant
for the kinetic mass and the energy shift, respectively.
Both have been calculated perturbatively by Davies and
Thacker~\cite{DT} and by Morningstar~\cite{MS}.
Since the precise form of their NRQCD action is slightly
different from ours, we performed the perturbative calculations for
our action.
Our results for the coefficient $A$ and $B$ in the
perturbative expansion
\begin{eqnarray}
  \label{eq:mass_renorm_2}
  E_0 &=& \alpha_V(q^*)A, \\
  \label{eq:mass_renorm_3}
  Z_m      &=& 1 + \alpha_V(q^*)B,
\end{eqnarray}
are summarized in Table~\ref{tab:selffac}.

\section{Simulation Details}
\label{sec:method}

Our task is to compute the ratios
$\langle{\cal O}_{\{L,R,N,S\}}^{lat}(1/a)\rangle
 /(8/3)\langle A_0^{lat}(1/a) \rangle^2$
using lattice NRQCD with the lattice spacing $a$.
In this section we describe our simulation method to obtain
them.

We performed the numerical simulation on a $16^3 \times 48$
lattice at $\beta$= 5.9, for which the inverse lattice
spacing fixed with the string tension is $1/a$= 1.64 GeV.
In the quenched approximation we use 250 gauge
configurations, each separated by 2,000 pseudo-heat bath sweeps.
For the light quark we use the $O(a)$-improved
action~\cite{SW} at $\kappa$=0.1350, 0.1365. 
The clover coefficient is set to be $c_{\rm sw}=1/u_0^3$,
where $u_0\equiv\langle P_{plaq} \rangle^{1/4}$=0.8734 at
$\beta=5.9$.
The critical $\kappa$ value is $\kappa_c=$0.1401, and
$\kappa_s$ corresponding to the strange quark mass
determined from the $K$ meson mass is $\kappa_s$=0.1385.

For the heavy quark and anti-quark we use the lattice NRQCD
action~\cite{NRQCD} with the tadpole improvement
$U_{\mu}\rightarrow U_{\mu}/u_0$.
The precise form of the action is the same as the one we used in the
previous work~\cite{Our}.
We use both of the $O(1/m_Q)$ and $O(1/m_Q^2)$ actions in parallel in
order to see the effect of the higher order contributions.
The heavy (anti-)quark field in the relativistic
four-component spinor form is constructed with the inverse 
Foldy-Wouthuysen-Tani (FWT) transformation defined at the
tadpole improved tree level as in Ref.~\cite{Our}.
The heavy quark masses and the stabilization parameters
are $(am_Q,n)$=(10.0,2), (5.0,2), (3.0,2), (2.6,2) and (2.1,3).
These parameters approximately cover a mass scale between $4m_b$ and
$m_b$.

We label the time axis of our lattice as $t=[-24,23]$.
The heavy quark and anti-quark propagators are created from a local
source located at the origin ($t$=0 on our lattice) and evolve into
opposite temporal directions.
The light quark propagator is also solved with the same source
location and with a Dirichlet boundary condition at $t=-24$ and
$t=23$.
The $B$ and $\overline{B}$ mesons are constructed with local sink
operators.
Thus we have the four-quark operators at the origin and extract the
matrix elements from the following three-point correlation function
\begin{equation}
  C^{(3)}_{X}(t_1,t_2)
  = \sum_{\vec{x}_1}\sum_{\vec{x}_2}
  \langle 0 | {A_0^{lat}}^{\dag}(t_1,\vec{x}_1) O_X^{lat}(0,\vec{0})
              {A_0^{lat}}^{\dag}(t_2,\vec{x}_2) |0 \rangle,
\end{equation}
where $X$ denotes $L$, $R$, $N$ or $S$. 
Because of a symmetry under parity transformation,
$C^{(3)}_{L}(t_1,t_2)$ and $C^{(3)}_{R}(t_1,t_2)$ should
exactly coincide in infinitely large statistics.
Therefore we explicitly average them before the fitting
procedure we describe below.

To obtain the ratios $B_X^{lat}(1/a)$ we also define the following
two-point functions
\be
  C^{(2)}(t_1)
  &=& \sum_{\vec{x}}
  \langle 0 | A_0^{lat}(t_1,\vec{x}) {A_0^{lat}}^{\dag}(0,\vec{0})| 0
  \rangle ,\\
  C^{(2)}(t_2)
  &=& \sum_{\vec{x}}
  \langle 0 | A_0^{lat}(0,\vec{0}) {A_0^{lat}}^{\dag}(t_2,\vec{x})| 0
  \rangle ,
\ee
and consider a ratio
\begin{equation}
  \label{eq:3pt_over_2pt}
  \frac{C^{(3)}_{X}(t_1,t_2)}{
    \frac{8}{3}C^{(2)}(t_1) C^{(2)}(t_2) }
  \longrightarrow 
  \hspace{-4.9ex}
  {\mbox{\scriptsize\raise 10pt \hbox{$|t_i|\gg 1$}}}
  \frac{\langle\overline{P^0} | {\cal O}_X^{lat}(1/a)
    | P^0\rangle}{
    \frac{8}{3}
    \langle\overline{P^0} |A_0^{lat}(1/a)| 0 \rangle
    \langle 0 |A_0^{lat}(1/a)| P^0\rangle }
  = B_X^{lat}(1/a),
\end{equation}
where $P^0$ denotes a heavy-light pseudo-scalar meson.
The ground state $P^0$ meson is achieved in the large
$|t_i|$ regime.
Although we use the local operator for the sinks at $t_1$ and
$t_2$, the ground state extraction is rather easier for
finite $am_Q$ than in the static approximation, since the
statistical error is much smaller for
NRQCD~\cite{Lepage91,Hashimoto}. 
This is another advantage of introducing the $1/m_Q$ correction.

The physical $B_B(m_b)$ is obtained by extrapolating and interpolating
each $B_X^{lat}(1/a)$ to the physical $B$ meson with $\kappa$ and
$m_Q$, respectively before combining them as
Eq.~(\ref{eq:matching_ratio}).
The final result for $B_B(m_b)$ may also be obtained by combining the
ratio of correlation functions before a constant fit.
Namely we use the relation
\begin{equation}
  \label{eq:combine-then-fit}
  B_P(m_b;t_1,t_2) = 
  \sum_{X=L,R,N,S} Z_{X/A^2}(m_b,1/a)
  \frac{C^{(3)}_{X}(t_1,t_2)}{
    \frac{8}{3}C^{(2)}(t_1) C^{(2)}(t_2) }
  \longrightarrow 
  \hspace{-4.9ex}
  {\mbox{\scriptsize\raise 10pt \hbox{$|t_i|\gg 1$}}}
  B_P(m_b).
\end{equation}
Since the statistical fluctuation in the individual
$B_X^{lat}(1/a)$ is correlated, the error is expected to be smaller
with this method (We use the jackknife method for error estimation).
Following Ref.~\cite{CDM} we refer to this method as the
``combine-then-fit'' method, while the usual one as
Eq.~(\ref{eq:3pt_over_2pt}) is called the ``fit-then-combine''
method in the rest of the paper. 

\section{Simulation Results}
\label{sec:results}

We describe the simulation results in this section.
The results are from the $O(1/m_Q)$ action
unless we specifically mention. 

\subsection{Heavy-light meson mass}
\label{sec:M_P}

The binding energy of the heavy-light meson is obtained from
a simultaneous fit of two two-point correlation functions.
The numerical results are listed in Table~\ref{tab:mass} for
each $am_Q$ and $\kappa$.
Extrapolation of the light quark mass to the strange quark
mass or to the chiral limit is performed assuming a linear
dependence in $1/\kappa$.

The meson mass is calculated using the perturbative
expression Eq.~(\ref{eq:mass_renorm_1}).
The results with $\alpha_V(\pi/a)$ and with $\alpha_V(1/a)$
are also given in Table \ref{tab:mass}.

\subsection{$B_X^{lat}(1/a)$ and $B_P(m_b)$}
\label{sec:B_X}

\begin{figure}
  \begin{center}
    \begin{tabular}{cc}
      \leavevmode\psfig{file=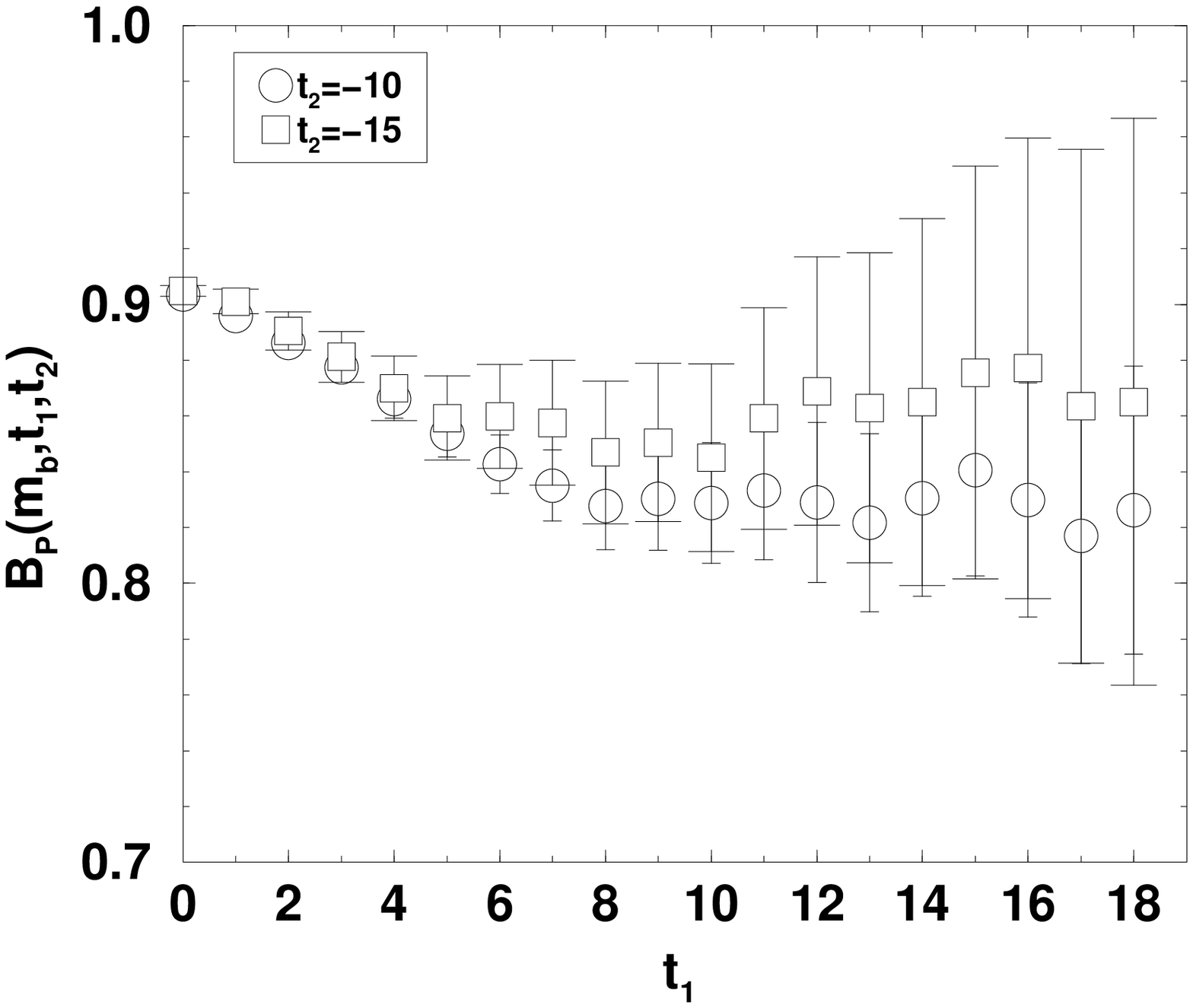,width=7.5cm}&
      \leavevmode\psfig{file=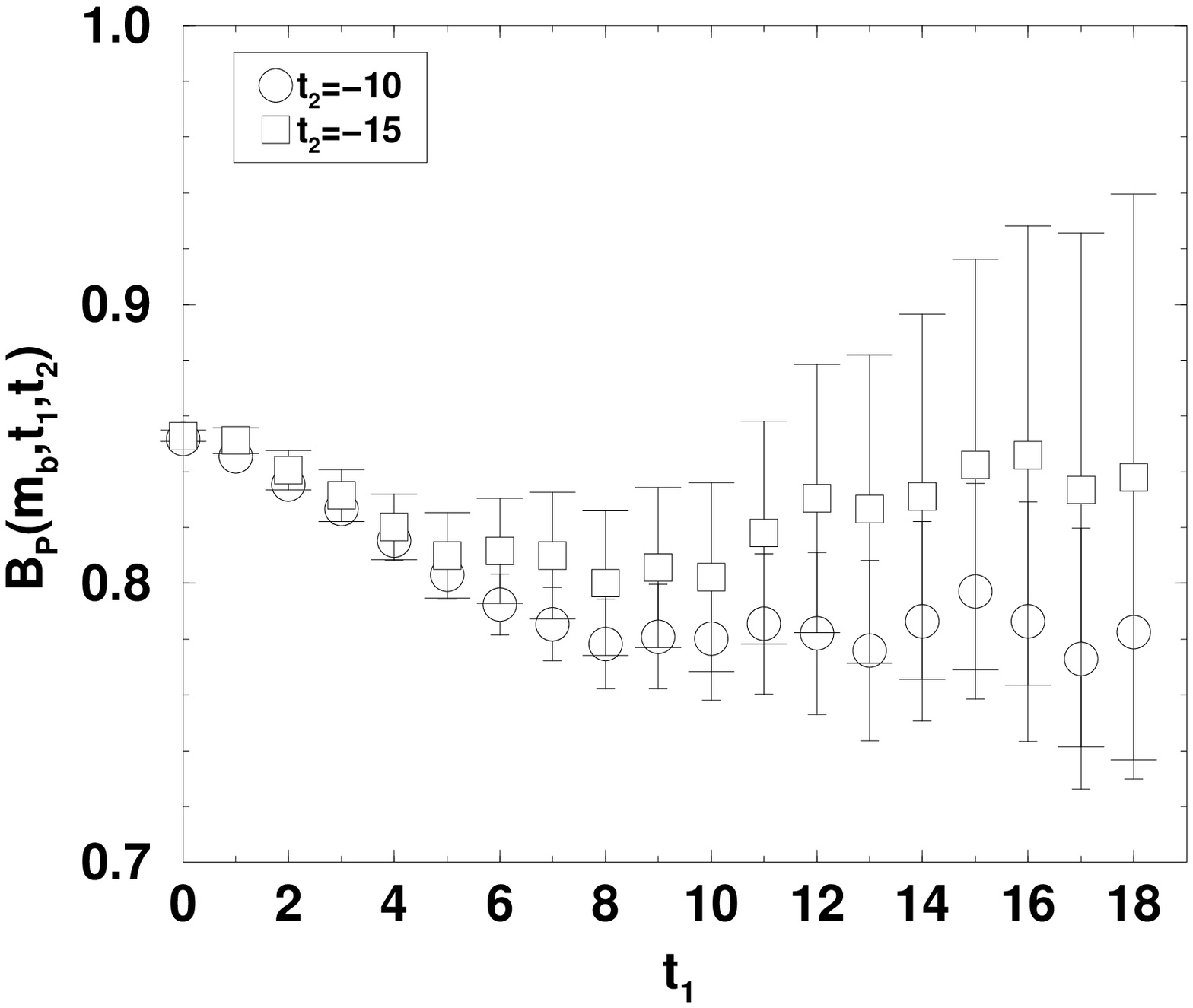,width=7.5cm}
    \end{tabular}
    \caption{$B_P(m_b;t_1,t_2)$ as a function of $t_1$,
      while $t_2$ is fixed at $-$10 (circles) or $-$15
      (squares).
      The heavy quark mass is $am_Q$=5.0 and $\kappa$=0.1365.
      The ``combine-then-fit'' method is used.
      In the perturbative matching, $\alpha_V(\pi/a)$ and
      $\alpha_V(1/a)$ are used in the left and right plot
      respectively. }
    \label{fig:B_B_5.0}
  \end{center}
\end{figure}

\begin{figure}
  \begin{center}
    \begin{tabular}{cc}
      \leavevmode\psfig{file=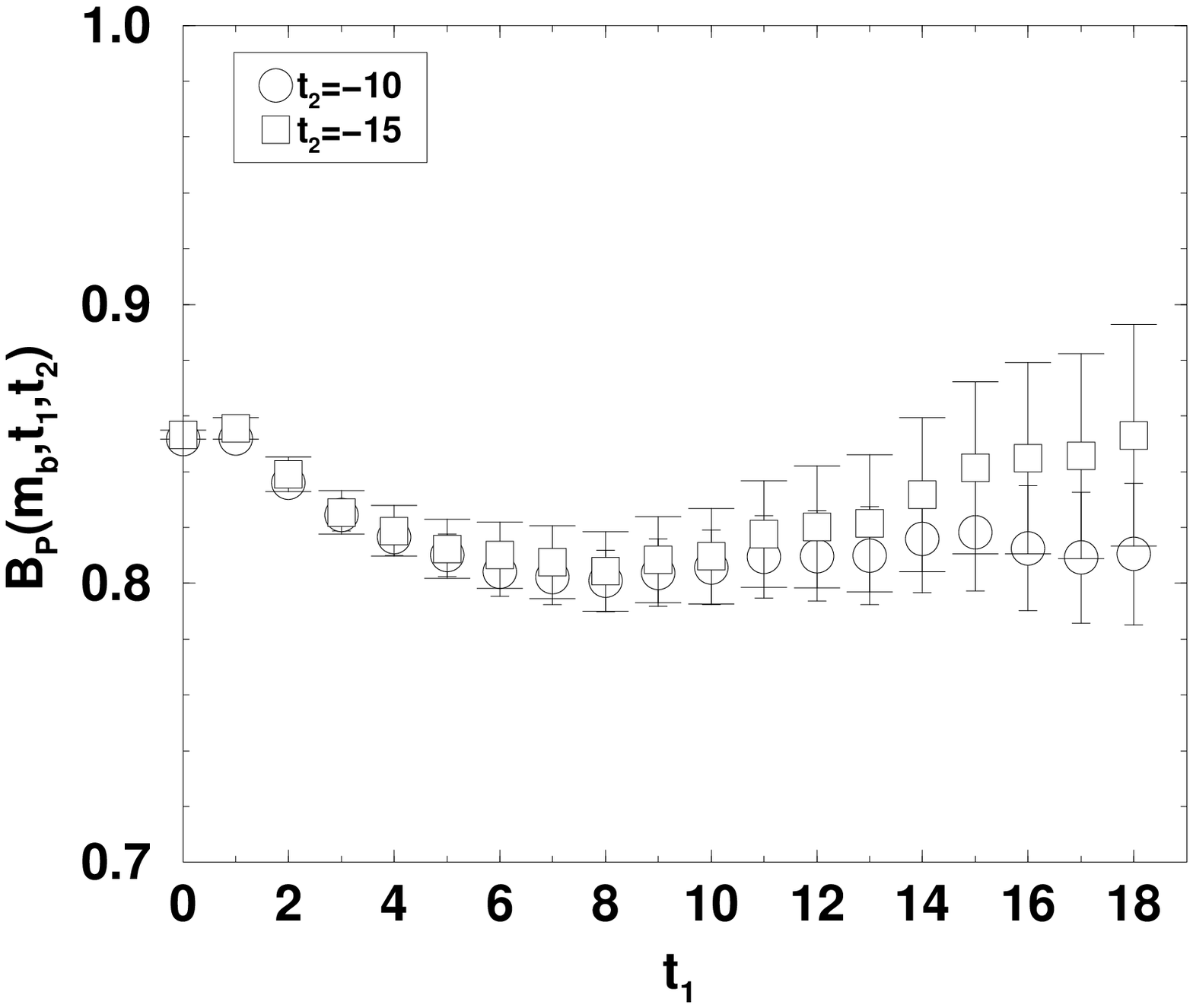,width=7.5cm}&
      \leavevmode\psfig{file=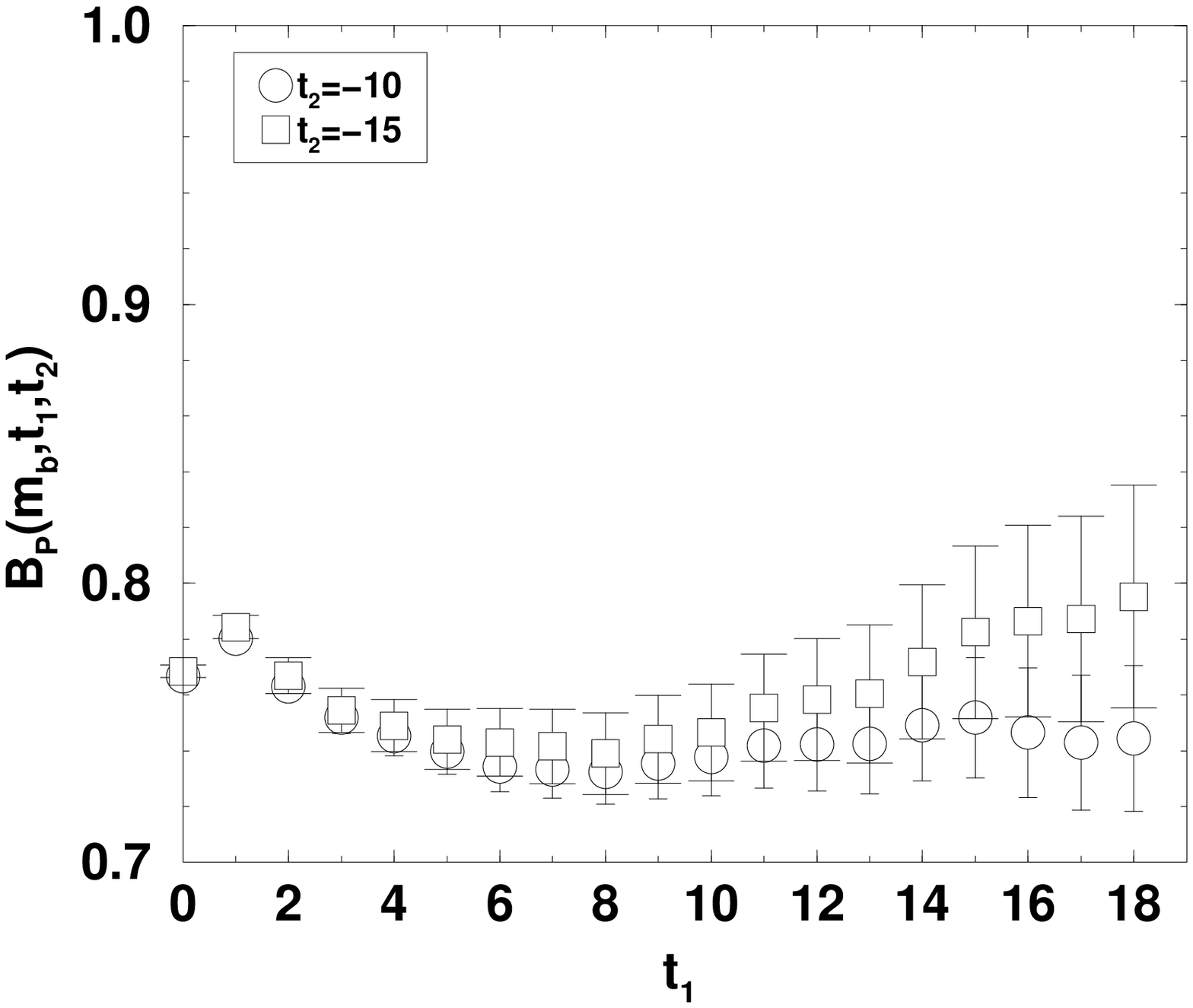,width=7.5cm}
    \end{tabular}
    \caption{Same as Figure \ref{fig:B_B_5.0} but at
      $am_Q$=2.6.} 
    \label{fig:B_B_2.6}
  \end{center}
\end{figure}

Figures \ref{fig:B_B_5.0} and \ref{fig:B_B_2.6} show the
$t_1$ dependence of $B_P(m_b;t_1,t_2)$ in the
``combine-then-fit'' method. 
The perturbative matching of the continuum and lattice
theory is done with the $V$-scheme coupling $\alpha_V(q^*)$ at
$q^*$=$\pi/a$ (left) and $1/a$ (right).
Their difference represents the effect of $O(\alpha_s^2)$.
The signal is rather noisier at $am_Q$=5.0 (Fig.~\ref{fig:B_B_5.0})
than at $am_Q$=2.6 (Fig.~\ref{fig:B_B_2.6}) from the same reason as
in the static limit\cite{Lepage91,Hashimoto}. 
But, still, a reasonably good signal is observed even for
large $am_Q$.
A plateau in the $t_1$ dependence is reached around $t_1$= 8
$\sim$ 11 for both $t_2$ = $-$10 and $-$15.
To be conservative we take $|t_1|$ as well as $|t_2|$ greater than
10 for the fitting region.
All data points $(t_1,t_2)$ in 10 $\leq|t_1|\leq$ 13 and
in 10 $\leq|t_2|\leq$ 13 are fitted with
constant to obtain the result for $B_P(m_b)$.
We confirm that except for the heaviest quark the results are stable
within about one standard deviation under a change of the fitting
region by at most two $t_i$ steps in the forward and backward direction.
The numerical results are listed in Table~\ref{tab:B_X}.

\begin{figure}
  \begin{center}
    \begin{tabular}{cc}
      \leavevmode\psfig{file=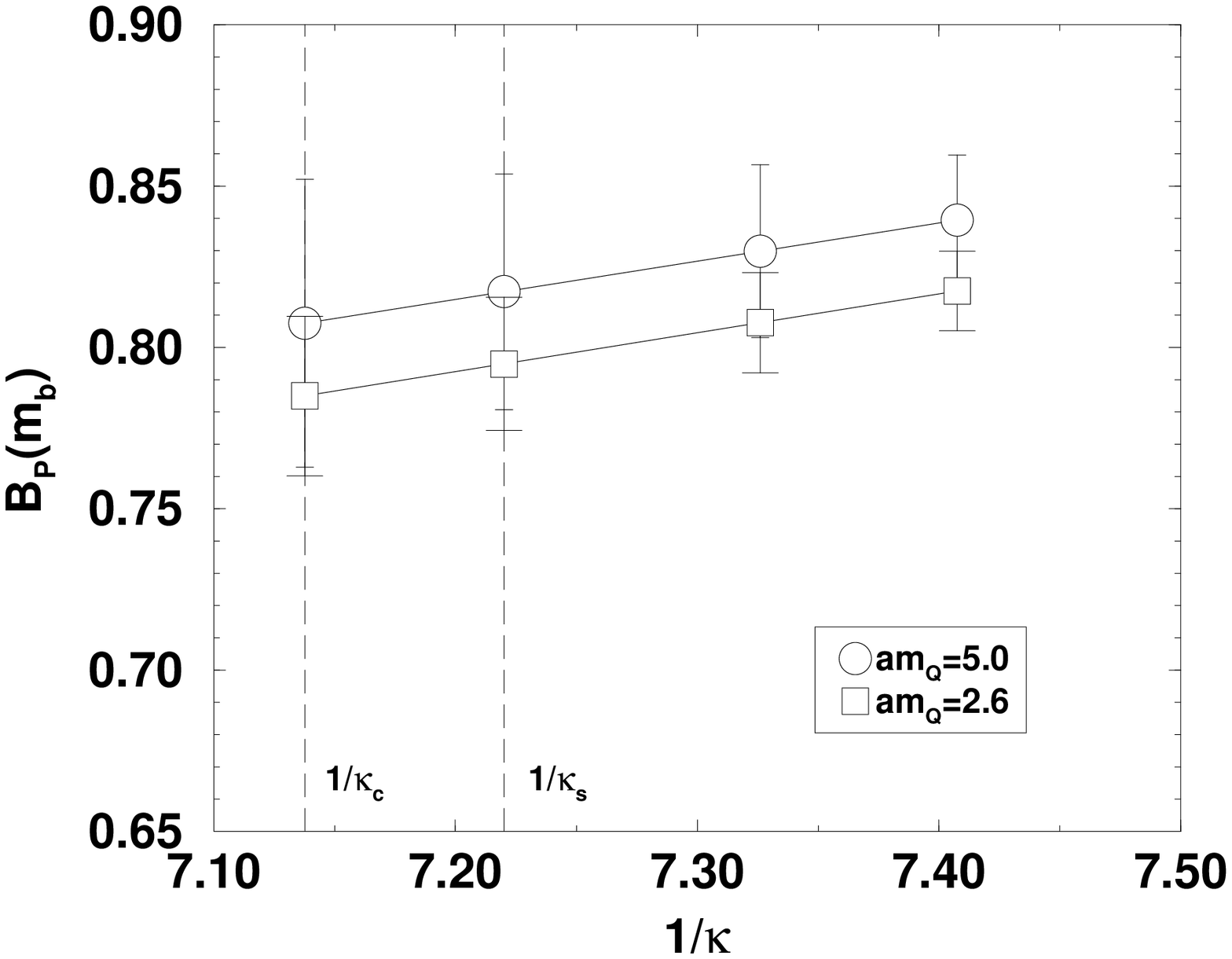,width=7.5cm}&
      \leavevmode\psfig{file=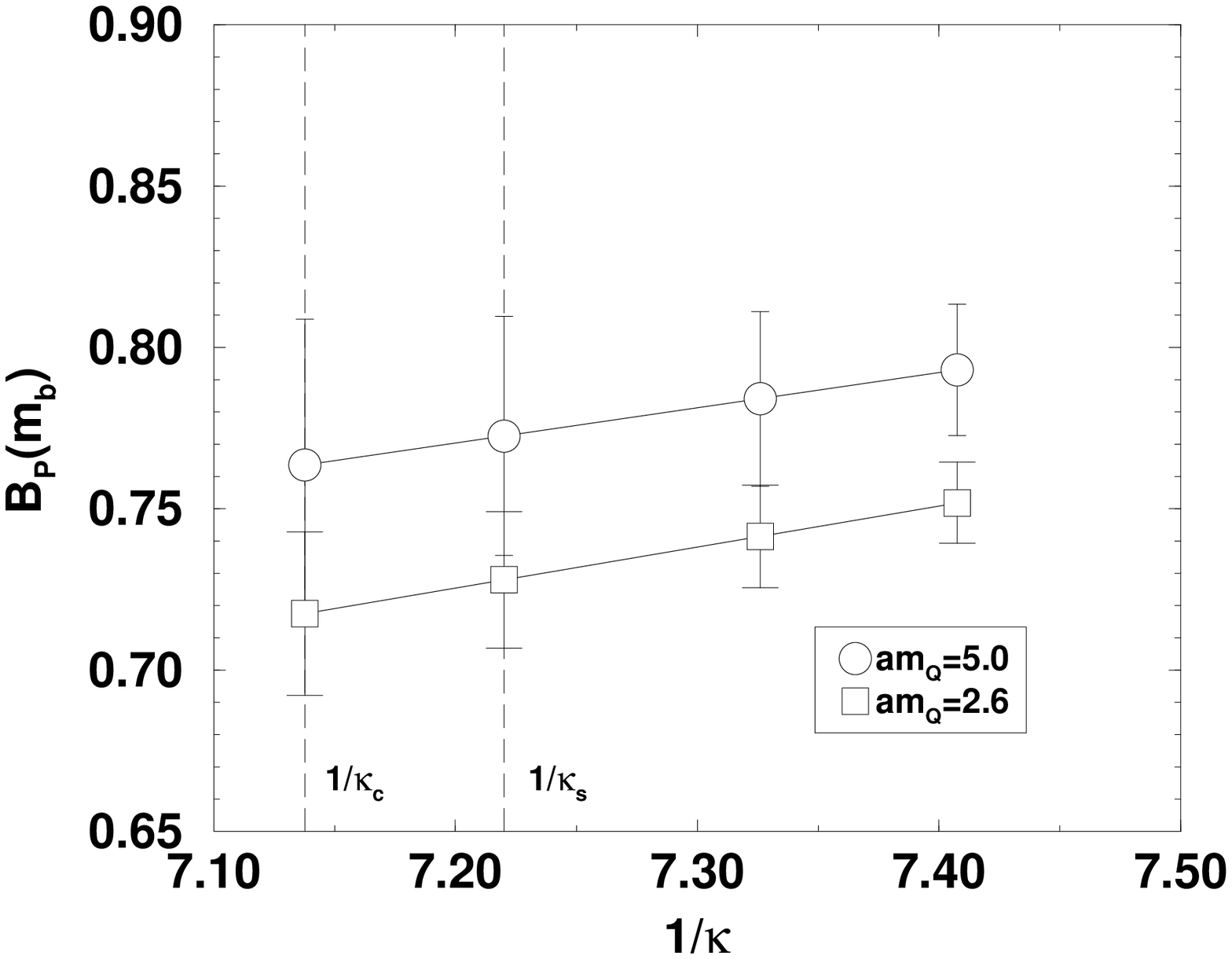,width=7.5cm}
    \end{tabular}
    \caption{Extrapolation of $B_P(m_b)$ to the strange and
      to the chiral limit.
      The heavy quark mass is $am_Q$=5.0 (circles) and 2.6
      (squares).
      Results with $q^*$=$\pi/a$ (left) and $1/a$ (right)
      are shown.}
    \label{fig:BBextrp}
  \end{center}
\end{figure}

The light quark mass ($1/\kappa$) dependence of $B_P(m_b)$
is presented in Fig.~\ref{fig:BBextrp}.
Since its dependence is quite modest, we assume a linear
dependence on $1/\kappa$ and extrapolate the results to the
strange quark mass and to the chiral limit as shown in the
plot.
Results of the extrapolation are also listed in Table~\ref{tab:B_X}.

\subsection{$1/M_P$ dependence}
\label{sec:1/M_P_dependence}

\begin{figure}
  \begin{center}
    \leavevmode\psfig{file=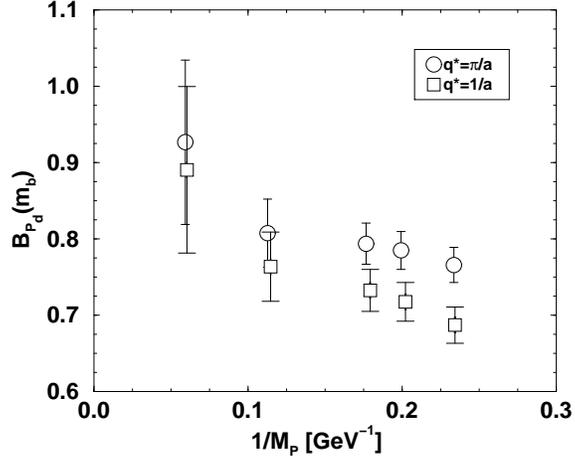,width=7.5cm}
    \caption{Inverse heavy-light meson mass dependence of
      $B_{P_d}(m_b)$ with $q^*$=$\pi/a$ (circles) and $1/a$
      (squares).} 
    \label{fig:BBmdep}
  \end{center}
\end{figure}

In Fig.~\ref{fig:BBmdep} we plot $B_P(m_b)$ in the chiral limit,
namely $B_{P_d}(m_b)$, as a function of $1/M_P$ in the physical unit.
We take $q^*$=$\pi/a$~(circles) and $1/a$~(squares) for the scale of
$\alpha_V$.
Regardless of the choice of the coupling, we observe a small
but non-zero negative slope in $1/M_P$, which supports the
previous results by Bernard, Blum and Soni~\cite{BBM} and by
Lellouch and Lin~\cite{LL} using the relativistic fermions.

To investigate the origin of the observed $1/M_P$
dependence, we look into the contributions of the individual
operators ${\cal O}_{\{L,R,N,S\}}^{lat}$ through the
``fit-then-combine'' method with the same fitting region as before.
We list the results for each $B_X^{lat}$ in Table~\ref{tab:B_X}. 
Figure~\ref{fig:ftc1} shows the $1/M_P$ dependence of
$B_L^{lat}(1/a)$(=$B_R^{lat}(1/a)$), $B_N^{lat}(1/a)$ and
$B_S^{lat}(1/a)$. 
While no significant $1/M_P$ dependence is observed in
$B_L^{lat}(1/a)$, $B_N^{lat}(1/a)$ and $B_S^{lat}(1/a)$ have strong
slope.
Since their sign is opposite and the sign of the matching factors
$Z_{N/A^2}(m_b;1/a)$ and $Z_{S/A^2}(m_b;1/a)$ (see
Table~\ref{tab:matchfac}) is the same, a partial cancellation
takes place giving a small negative slope for $B_{P_d}(m_b)$.

We also make a comparison of the results of the $1/m_Q$
action~(circles) with those of the $1/m_Q^2$ action~(triangles)
in Fig.~\ref{fig:ftc1}.
There is a small difference between the two results in
$B_N^{lat}(1/a)$ and in $B_S^{lat}(1/a)$ toward large $1/M_P$
($1/M_P\geq$ 0.2 GeV$^{-1}$), which is consistent with our
expectation that the difference is an $O(\Lambda_{QCD}/m_Q)^2$
effect.

Previous results in the static approximation
by Ewing \textit{et al.}~(diamond)~\cite{UK}, Gimen\'ez and
Martinelli~(triangle)~\cite{GM} and Christensen, Draper and
McNeile~(circle)~\cite{CDM} are plotted with filled symbols
at $1/M_P=0$.
Although the $\beta$ value and the light quark action
employed (the $O(a)$-improved action is used in Refs.~\cite{UK,GM} and
unimproved action in Ref.~\cite{CDM}) are different, all the results
are in good agreement with each other.
A quadratic extrapolation~(dashed line) using our $1/m_Q^2$ NRQCD result
also does agree with these previous static results.

\subsection{Result for $B_B(m_b)$}
\label{sec:Result_for_B_B}

\begin{figure}
  \begin{center}
    \leavevmode\psfig{file=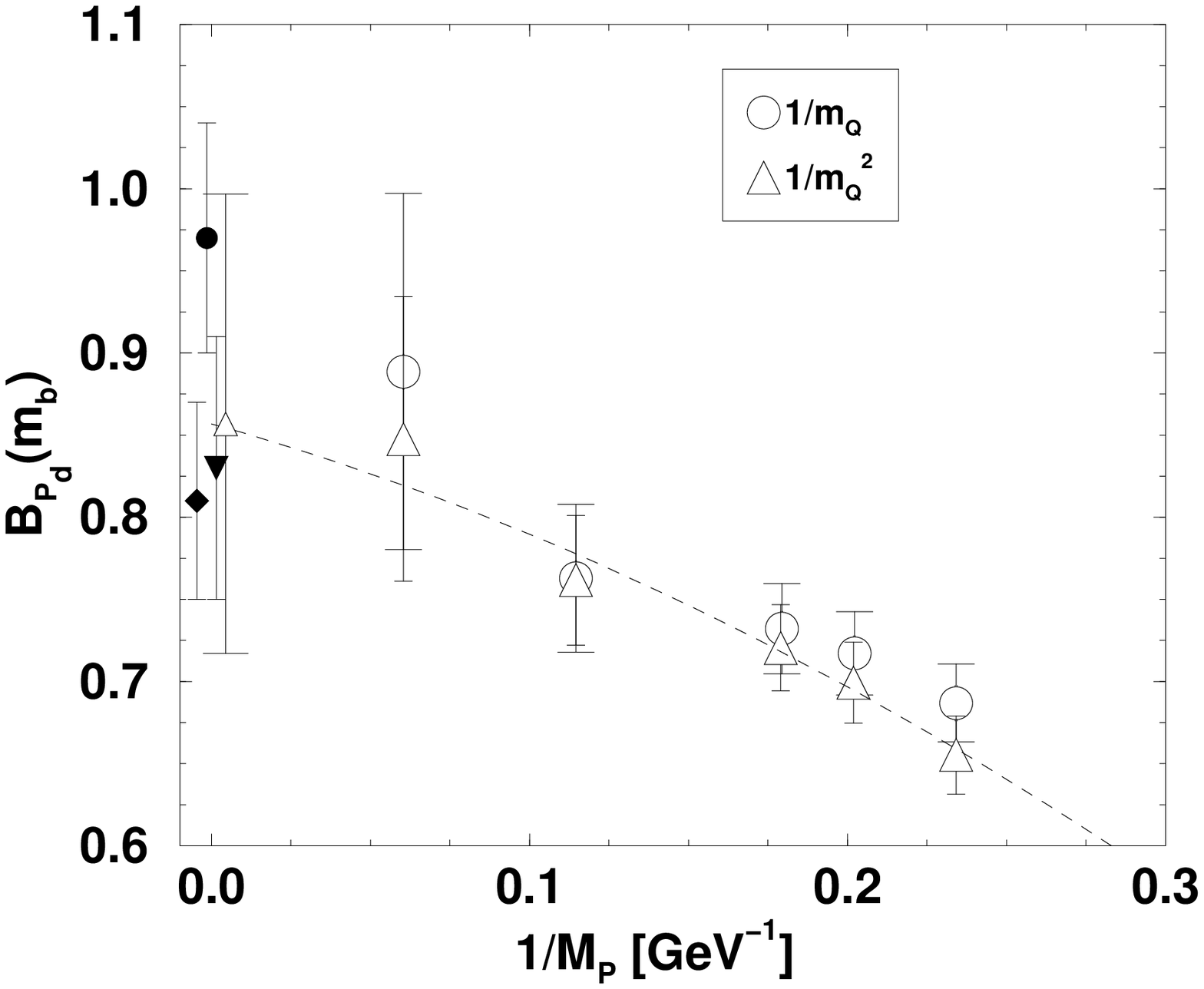,width=7.5cm}
    \caption{
      Comparison of our NRQCD data (open symbols)
      with the static ones (filled symbols).
      }
    \label{fig:BBcomp}
  \end{center}
\end{figure}

Combining the data for $B_X^{lat}(1/a)$ discussed above, we obtain
$B_{P_d}(m_b)$ with the ``fit-then-combine'' method.
We confirm that the difference in numerical results from both methods are
completely negligible.
Figure~\ref{fig:BBcomp} shows the results of ``fit-then-combine''
method using $\alpha_V(1/a)$ with both actions.
The comparison with the static results is also made in this
plot, where only the statistical error in each calculation is
considered and the same matching procedure as ours are applied.
We again observe a consistent result.

Interpolating the above NRQCD results to the physical $B$
meson mass, we obtain the physical $B_{B_d}(m_b)$
\begin{equation}
  \label{eq:result_1}
  B_{B_d}(m_b) = 
  \left\{
    \begin{array}{c}
      0.78(3) \quad (q^*=\pi/a)\\
      0.72(3) \quad (q^*=1/a)
    \end{array}      
  \right.
\end{equation}
for the $O(1/m_Q)$ action, and 
\begin{equation}
  \label{result_2}
  B_{B_d}(m_b) =
  \left\{
    \begin{array}{c}
      0.78(2) \quad (q^*=\pi/a)\\
      0.71(3) \quad (q^*=1/a)
    \end{array}      
  \right.
\end{equation}
for the $O(1/m_Q^2)$ action.
The quoted error is statistical only.
For the ratio of $B_{B_s}/B_{B_d}$, we obtain
$B_{B_s}/B_{B_d}$ = 1.01(1) for $q^*=\pi/a$ and
$B_{B_s}/B_{B_d}$ = 1.02(1) for $q^*=1/a$
from both actions.

\section{Discussion}
\label{sec:VSA}

\begin{figure}
  \begin{center}
    \leavevmode\psfig{file=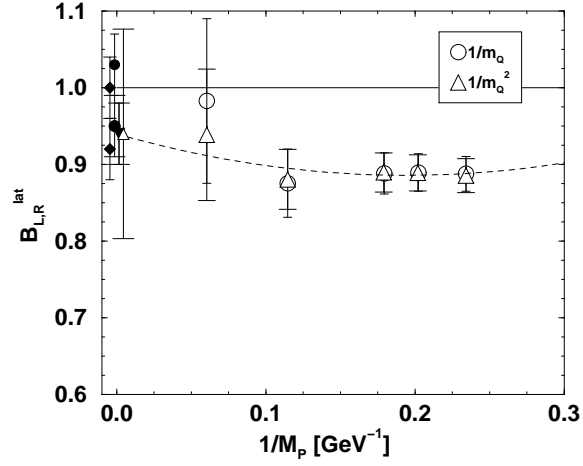,width=7.5cm}\\(a)\\
    \leavevmode\psfig{file=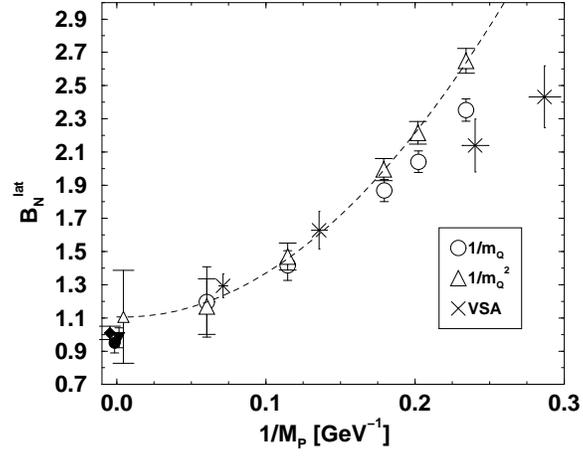,width=7.5cm}\\(b)\\
    \leavevmode\psfig{file=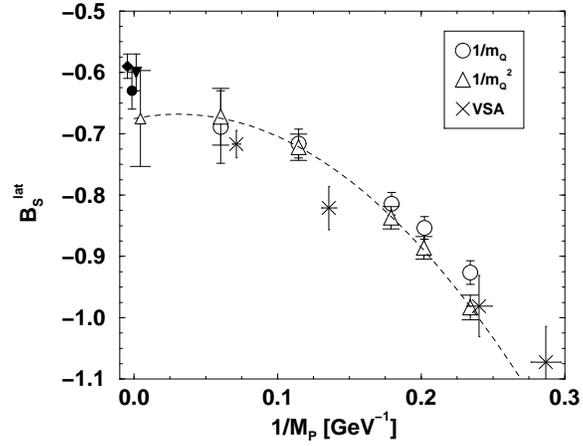,width=7.5cm}\\(c)\\
    \caption{Inverse heavy-light meson mass dependence of
      (a) $B_{L,R}^{\rm lat}$, (b) $B_N^{\rm lat}$, 
      (c) $B_S^{\rm lat}$. For what symbols and line denote, see
      text.}
    \label{fig:ftc1}
  \end{center}
\end{figure}

The strong $1/M_P$ dependence in $B_X^{lat}(1/a)$ observed in
Fig.~\ref{fig:ftc1} can be roughly understood using the
vacuum saturation approximation (VSA) on the lattice as explained
below.
Here it should be noted that the terminology of VSA we use here does
not immediately mean $B_B(m_b)=1$.

The VSA for $B_{L,R}^{lat}$ is unity by construction.
This is true even for finite $1/M_P$, and its prediction is
shown by a straight line in Fig.~\ref{fig:ftc1}(a). 
The NRQCD data is located slightly below the line ($\sim$0.9),
but the mass dependence is well reproduced by the VSA. 

For $B_{N}^{lat}$ and $B_{S}^{lat}$, we require a little
algebra to explain their mass dependence under the VSA.
Using the Fierz transformation and inserting the vacuum, we obtain
\begin{eqnarray}
  \label{eq:ONvsa}
  \langle\overline{P^0}|{\cal O}_N^{lat}|P^0\rangle
  & = &
  - \frac{8}{3}
  \langle\overline{P^0} | \bar{b}\gamma_{\mu}\gamma_5 q
  | 0 \rangle
  \langle 0 | \bar{b}\gamma_{\mu}\gamma_5 q 
  | P^0 \rangle
  - \frac{16}{3}
  \langle\overline{P^0}|\bar{b}\gamma_5 q| 0 \rangle
  \langle 0 | \bar{b}\gamma_5 q| P^0 \rangle, \\
  \label{eq:OSvsa}
  \langle\overline{P^0}|\Op_S^{lat}|P^0\rangle
  & = &
  \frac{5}{3}
  \langle\overline{P^0}|\bar{b}\gamma_5 q| 0 \rangle
  \langle 0 | \bar{b}\gamma_5 q| P^0 \rangle,
\end{eqnarray}
where $|P^0\rangle$ denotes a heavy-light pseudo-scalar meson
at rest, and 
$\langle 0 | \bar{b}\gamma_{\mu}\gamma_5 q | P^0 \rangle$ is
related to the pseudo-scalar decay constant
\begin{equation}
  \langle\overline{P^0}|A_0^{lat}(1/a)| 0 \rangle
  = \langle 0 |A_0^{lat}(1/a)| P^0 \rangle
  = f_P M_P.
\end{equation}

Let us now consider a decomposition of the $b$-quark field
$\bar{b}$ into the two-component non-relativistic quark
$Q^{\dagger}$ and anti-quark $\chi$ fields.
Up to $O(1/m_{Q}^2)$ we have 
\begin{eqnarray}
  \bar{b}\gamma_5 q & = &
  ( Q^\dag\quad 0 )
  \left( 1 + \frac{\gam\cdot\nabd}{2m_Q} \right)
  \gamma_5 q -
  ( 0\quad \chi )
  \left( 1 + \frac{\gam\cdot\nabd}{2m_Q} \right)
  \gamma_5 q, \\
  \bar{b}\gamma_0\gamma_5 q & = &
  ( Q^\dag\quad 0 )
  \left( 1 - \frac{\gam\cdot\nabd}{2m_Q} \right)
  \gamma_5 q +
  ( 0\quad \chi )
  \left( 1 - \frac{\gam\cdot\nabd}{2m_Q} \right)
  \gamma_5 q,
\end{eqnarray}
and then
\begin{eqnarray}
  \langle \overline{P^0} | \bar{b}\gamma_5 q |0\rangle
  & = &
  \langle \overline{P^0} | ( Q^\dag\quad 0 )
  \left( 1 + \frac{\gam\cdot\nabd}{2m_Q} \right)
  \gamma_5 q |0\rangle \nonumber \\
  & = &
  \langle \overline{P^0} | \bar{b}\gamma_0
  \gamma_5 q |0\rangle 
  + 2 \langle \overline{P^0} | ( Q^\dag\quad 0 ) 
  \frac{\gam\cdot\nabd}{2m_Q} \gamma_5 q |0\rangle, \\
  \langle 0| \bar{b}\gamma_5 q |P^0\rangle
  & = & 
  - \langle 0| ( 0\quad \chi )
  \left( 1 + \frac{\gam\cdot\nabd}{2m_Q} \right)
  \gamma_5 q |P^0\rangle \nonumber \\
  & = &
  - \langle 0| \bar{b}\gamma_0\gamma_5 q |P^0\rangle
  + 2 \langle 0| ( 0\quad \chi )
  \frac{\gam\cdot\nabd}{2m_Q} \gamma_5 q |P^0\rangle.
\end{eqnarray}
By defining $\delta f_P$ as 
\begin{equation}
  - \langle \overline{P^0} | ( Q^\dag\quad 0 ) 
  \frac{\gam\cdot\nabd}{2m_Q} \gamma_5 q |0\rangle
  = - \langle 0| ( 0\quad \chi )
  \frac{\gam\cdot\nabd}{2m_Q} \gamma_5 q |P^0\rangle
  \equiv \delta f_P M_P,
\end{equation}
we obtain
\begin{eqnarray}
  \langle\overline{P^0}|{\cal O}_N^{lat}|P^0\rangle
  & = &
  \frac{8}{3} f_P^2 M_P^2
  \left( 1 - 8 \frac{\delta f_P}{f_P} \right),\\
  \langle\overline{P^0}|{\cal O}_S^{lat}|P^0\rangle
  & = &
  - \frac{5}{3} f_P^2 M_P^2
  \left( 1 - 4 \frac{\delta f_P}{f_P} \right).
\end{eqnarray}
In our previous work~\cite{Our} we denoted 
$\delta f_P$ as $\delta f_P^{(2)}$.

Thus the VSA for $B_N^{lat}$ and for $B_S^{lat}$ read
\begin{eqnarray}
  \label{eq:VSA_B_N}
  B_N^{lat\rm (VSA)} & = &
  1 - 8 \frac{\delta f_P}{f_P}, \\
  \label{eq:VSA_B_S}
  B_S^{lat\rm (VSA)} & = &
  - \frac{5}{8}
  \left[ 1 - 4 \frac{\delta f_P}{f_P} \right],
\end{eqnarray}
neglecting the higher order contribution of order $1/m_Q^2$. 
Results for $\delta f_P/f_P$ is available at $\beta$=5.8 in
Ref.~\cite{Our}.
We plot them in Fig.~\ref{fig:ftc1}(b) and \ref{fig:ftc1}(c)
by crosses, which show a qualitative agreement with the
measured values.

Di~Pierro and Sachrajda~\cite{PS} pointed out that the value
of several $B$-parameter-like matrix elements of the $B$ meson
is explained by the VSA surprisingly well in the static limit.
Here we find that the $1/m_Q$ dependence of $B_X^{lat}(1/a)$
can also be reproduced qualitatively.
This result suggests that the vacuum saturation is a reasonable
qualitative picture for the heavy-light meson.
It does not mean, however, that the VSA works quantitatively for
$B_B(\mu)$, and careful lattice studies are necessary for precise
calculation of the $B$-parameters.

\section{Remaining uncertainties and the final result}
\label{sec:uncertainty}

To estimate the systematic uncertainties in lattice
calculations is a difficult task.
In our case it is even more true, since we have a simulation
result only at a single $\beta$ value.
However we attempt to do it, giving a dimension counting of
missing contributions.

The following sources of systematic errors are possible:
\begin{itemize}
\item discretization error:
  Both of the heavy and light quark actions are
  $O(a)$-improved at tree level, and there is no
  discretization error of $O(a\Lambda_{QCD})$.
  The leading error is of $O(a^2\Lambda_{QCD}^2)$ and of
  $O(\alpha_s a\Lambda_{QCD})$.
  The second one is from the missing one-loop perturbative
  correction in the $O(a)$-improvement~(its matching coefficient has
  been already obtained in Ref.~\cite{IOY}).
  We naively estimate the size of them to be
  $O(a^2\Lambda_{QCD}^2)\sim 
   O(a\Lambda_{QCD}\alpha_s)\sim$ 5\%, assuming
  $1/a\sim$1.6 GeV, $\Lambda_{QCD}\sim$300 MeV and
  $\alpha_s\sim$0.3. 
\item perturbative error:
  The operator matching of the continuum and lattice
  $\Delta B$=2 operators are done at one-loop level.
  Thus the $O(\alpha_s^2)$ correction is another source of error.
  In addition, we use the one-loop coefficient for the
  static lattice action, though our simulation has been
  done with the NRQCD action.
  The error in this mismatch is as large as
  $O(\alpha_s/(am_Q))$ and $O(\alpha_s\Lambda_{QCD}/m_Q)$.
  The size of these contributions is 
  $O(\alpha_s^2)$ $\sim$ $O(\alpha_s/(am_Q))$ $\sim$ 10\% and
  $O(\alpha_s\Lambda_{QCD}/m_Q)$ $\sim$ 2\%.
\item relativistic error:
  Since we have performed a set of simulations with the
  $O(1/m_Q)$ action and the $O(1/m_Q^2)$ action, we can
  estimate the error in the truncation of the
  non-relativistic expansion.
  As we have shown, the difference between the results with
  $O(1/m_Q)$ and $O(1/m_Q^2)$ is small ($\sim$ 2\%) around
  the $B$ meson mass.
  Then the higher order ($O(1/m_Q^3)$) effect is negligible.
\item chiral extrapolation:
  We have only two light quark $\kappa$ values.
  Then the linear behavior in the chiral extrapolation is
  nothing but an assumption.
  Although the light quark mass dependence is small and the
  assumption is a reasonable one, we conservatively estimate
  the error from the difference between the data at our
  lightest $\kappa$ value ($\kappa$=0.1365) and $\kappa_c$.
  It leads 3\% for $B_{B_d}(m_b)$.
\item quenching error:
  All results are obtained in the quenched approximation.
  Study of the sea quark effect is left for future work.
\end{itemize}

Taking them into account, we obtain the following values as
our final results from the quenched lattice,
\begin{eqnarray}
  B_{B_d}(m_b) & = & 
  0.75(3)(12),\nonumber \\
  \frac{B_{B_s}}{B_{B_d}} & = & 1.01(1)(3),\nonumber
\end{eqnarray}
where the first error is statistical and the second a sum of
all systematic errors in quadrature. 
In estimating the error of the ratio
$B_{B_s}/B_{B_d}$ we consider the error from chiral extrapolation
only, assuming that other uncertainties cancel in the ratio.
The above result is related to the scale invariant $B$-parameter
$\hat{B}_{B_d}$ as 
\begin{equation}
  \hat{B}_{B_d} =
  \left\{\begin{array}{l}\displaystyle
      \left[\alpha_s(m_b)\right]^{-6/23}B_{B_d}(m_b)
      = 1.12(4)(18)\\ \displaystyle
      \left[\alpha_s(m_b)\right]^{-6/23}
      \left[ 1 + \frac{\alpha_s(m_b)}{4\pi}J_5 \right]
      B_{B_d}(m_b)
      = 1.15(5)(18)
    \end{array}      
  \right., \nonumber
\end{equation}
using the leading and next-to-leading formula, respectively,
where we use $\Lambda_{QCD}^{(5)}$=0.237 GeV and the two-loop
$\beta$-function.

\section{Conclusion}
\label{sec:conclusion}

In this paper we investigate the $O(\Lambda_{QCD}/m_Q)$ and 
$O(\Lambda_{QCD}^2/m_Q^2)$ effects on the $B$-parameter.
We find that there is no significant mass dependence in the
leading operator contribution $B_L^{lat}(1/a)$, while the
mixing contributions $B_N^{lat}(1/a)$ and $B_S^{lat}(1/a)$
have large $O(\Lambda_{QCD}/m_Q)$ corrections.
The $O(\Lambda_{QCD}^2/m_Q^2)$ correction for each $B_X^{lat}(1/a)$
is, however, reasonably small for the $B$ meson as we naively
expected.
The observed $1/m_Q$ dependence is qualitatively
understood using the vacuum saturation approximation for
the lattice matrix elements.

The lattice NRQCD calculation predicts the small but non-zero
negative slope in the mass dependence of $B_P(m_b)$ and about 10\%
reduction from static limit to the physical $B$ meson.
In the present analysis, we combine lattice simulation for finite
heavy quark mass with the mass independent matching coefficients
determined in the static limit.
The dominant uncertainty is, therefore, arising from the finite mass
effects in the perturbative matching coefficients. 
For more complete understanding of the $1/m_Q$ dependence, matching
coefficients with the finite heavy quark mass are necessary.

\section*{Acknowledgment}
We would like to thank S. Tominaga for useful discussion.
Numerical calculations have been done on Paragon XP/S at
Institute for Numerical Simulations and Applied Mathematics
in Hiroshima University.
We are grateful to S. Hioki for allowing us to use his program
to generate gauge configurations.
T.O. is supported by the Grants-in-Aid
of the Ministry of Education (No. 10740125).
H.M. would like to thank the JSPS
for Young Scientists for a research fellowship.

\appendix
\section{}
\label{sec:appendix}

In this appendix, we compare our perturbative matching by simple
one-loop formula with the renormalization group (RG) improved ones
used in the previous static calculations~\cite{UK,GM,CDM}.
Since the matching procedure for determining the RG improved
coefficients is given in Refs.~\cite{UK,GM,CDM} in detail (see also
Refs.~\cite{Gimenez,CFG,Buchalla}), we just show the results
appropriate for our actions and definition of operator.

Considering the matching of a ratio
$B_B(m_b)=\langle{\cal O}_L(m_b)\rangle
 /(8/3)\langle A_0 \rangle^2$ again as in
section~\ref{sec:pertmat}, the RG improved versions of
$Z_{X/A^2}(m_b;1/a)$ are as follows,
\begin{eqnarray}
  \label{eq:coeff_L_app}
   Z_{L/A^2}(m_b;1/a) & = & Z_L^{cont} 
     \left( 1 + \frac{\alpha_s}{4\pi} (D_L-2D_A) \right), \\
  \label{eq:coeff_R_app}
   Z_{R/A^2}(m_b;1/a) & = & Z_L^{cont}
     \times \frac{\alpha_s}{4\pi} D_R, \\
  \label{eq:coeff_N_app}
   Z_{N/A^2}(m_b;1/a) & = & Z_L^{cont}
     \times \frac{\alpha_s}{4\pi} D_N, \\
  \label{eq:coeff_S_app}
   Z_{S/A^2}(m_b;1/a) & = & Z_S^{cont},
\end{eqnarray}
and
\begin{eqnarray}
  \label{eq:coeff_cont_L_app}
  Z_L^{cont} & = & 
    \left( 1 + \frac{\alpha_s(m_b)}{4\pi}(-\frac{26}{3})
    \right)
    \left( 1 + \frac{\alpha_s(1/a)-\alpha_s(m_b)}{4\pi}
               (0.043) \right) \nonumber \\
    & & + \frac{\alpha_s(m_b)}{4\pi} (-8)
    \left(
      \left(\frac{\alpha_s(m_b)}{\alpha_s(1/a)}\right)^{8/25}
      - 1 \right) \frac{1}{4}, \\
  \label{eq:coeff_cont_S_app}
  Z_S^{cont} & = & \frac{\alpha_s(m_b)}{4\pi} (-8)
    \left(\frac{\alpha_s(m_b)}{\alpha_s(1/a)}\right)^{8/25}.
\end{eqnarray}
Numerical values of $Z_{\{L,R,N,S\}/A^2}(m_b;1/a)$ are given in
Table~\ref{tab:matchfac} together with those by the simple one-loop
formula, where we use the V-scheme coupling~\cite{LM} as $\alpha_s$
appearing in Eqs.~(\ref{eq:coeff_L_app})-(\ref{eq:coeff_S_app}) while
the couplings in Eqs.~(\ref{eq:coeff_cont_L_app}) and
(\ref{eq:coeff_cont_S_app}) are defined in the continuum
$\overline{\rm MS}$ scheme with
$\Lambda_{\overline{\rm MS}}^{(4)}$=344 MeV, which corresponds to
$\Lambda_{\overline{\rm MS}}^{(5)}$= 237 MeV.

Now assuming each $B_X^{lat}$ is of $O(1)$, the dominant effects of
resummation arise from $Z_L/A^2$ and $Z_S/A^2$.
Since, however, its difference is at most 5\% level and the effects
from $Z_L/A^2$ and $Z_S/A^2$ are destructive, the total effect amount
to less than 3\%.
To be specific, using our data extrapolated to the static limit (see
Table~\ref{tab:B_X}) to calculate $B_B^{stat}(m_b)$, we obtain the
results tabulated in Table~\ref{tab:statB} from the two matching
procedures.
In this case the effect of resummation is almost negligible.

\begin{table}
  \begin{center}
    \begin{tabular}{cc|ccccc}
      &$q^*$   & $\alpha_V(q^*)$
        & $Z_{L/A^2}$ & $Z_{N/A^2}$ & $Z_{R/A^2}$ & $Z_{S/A^2}$\\
      \hline
      One-loop
      & $\pi/a$ & 0.164 & 0.973 & $-$0.080 & $-$0.007 & $-$0.104 \\
      & $1/a$   & 0.270 & 0.956 & $-$0.132 & $-$0.011 & $-$0.172 \\
      \hline
      RG improved
      & $\pi/a$ & 0.164 & 0.930 & $-$0.069 & $-$0.006 & $-$0.118 \\
      & $1/a$   & 0.270 & 0.978 & $-$0.113 & $-$0.010 & $-$0.118 \\
    \end{tabular}
    \caption{Matching factors at $\beta$=5.9 by the two different
    procedures.}
    \label{tab:matchfac}
  \end{center}
\end{table}
\begin{table}
  \begin{center}
    \begin{tabular}{c|ccccc}
      $a m_Q$ &     10.0 &   5.0 &   3.0 &   2.6 &   2.1 \\ 
      \hline
      $A$     &    1.011 & 0.946 & 0.855 & 0.819 & 0.754 \\
      $B$     & $-$0.075 & 0.018 & 0.119 & 0.152 & 0.329 \\
    \end{tabular}
    \caption{One-loop coefficients for the self-energy.} 
    \label{tab:selffac}
  \end{center}
\end{table}

\begin{table}
  \begin{center}
    \begin{tabular}{c|ccccc}
      $a m_Q$        &   10.0 &   5.0 &   3.0 &   2.6 &   2.1 \\ 
      \hline
      $\kappa$=0.1350 &
        0.691(10) & 0.675(4) & 0.664(3) & 0.660(3) & 0.653(2) \\
      $\kappa$=0.1365 &
        0.655(12) & 0.636(5) & 0.625(3) & 0.620(3) & 0.613(3) \\
      \hline
      $\kappa_s$=0.1385 &
        0.608(15) & 0.586(7) & 0.574(4) & 0.569(4) & 0.560(3) \\
      $\kappa_c$=0.1401 &
        0.571(18) & 0.547(8) & 0.534(5) & 0.529(4) & 0.520(4) \\
      \hline
      $M_P$[GeV]($q^*$=$\pi/a$) &
        16.864(30) & 8.867(13) & 5.662(8) & 5.018(7) & 4.279(7)\\
      $M_P$[GeV]($q^*$=$1/a$) &
        16.558(30) & 8.719(13) & 5.576(8) & 4.944(7) & 4.268(7)
    \end{tabular}
    \caption{The numbers in second and third lines are the
      binding energies obtained from the simultaneous
      fits. Those in fourth and fifth lines are the results
      of linear extrapolations to the strange and chiral
      limit. The numbers in last two rows are the
      corresponding physical meson masses with $q^*$=$\pi/a$
      and $1/a$.} 
    \label{tab:mass}
  \end{center}
\end{table}
\begin{table}
  \begin{center}
    \begin{tabular}{c|c|ccccc}
      $\kappa$ & $a m_Q$ &
      $B_L^{lat}(a)$ & $B_N^{lat}(a)$ & $B_S^{lat}(a)$ &
      $B_P(m_b)$ & $B_P(m_b)$ \\ 
      & & (=$B_R^{lat}(a)$) & & &
      with $\alpha_V(\pi/a)$ & with $\alpha_V(  1/a)$ \\ 
      \hline
      $\kappa$=0.1350
      &10.0 &
      0.94(5) & 1.27(8) & $-$0.69(3) & 0.88(4) & 0.84(5)\\
      & 5.0 &
      0.91(2) & 1.47(4) & $-$0.73(1) & 0.84(2) & 0.79(2)\\
      & 3.0 &
      0.92(1) & 1.85(3) & $-$0.82(1) & 0.83(1) & 0.77(1)\\
      & 2.6 &
      0.92(1) & 2.00(3) & $-$0.86(1) & 0.82(1) & 0.75(1)\\
      & 2.1 &
      0.92(1) & 2.27(4) & $-$0.93(1) & 0.80(1) & 0.72(1)\\
      \hline
      $\kappa$=0.1365
      &10.0 &
      0.95(6) & 1.25(12) & $-$0.69(4) & 0.89(6) & 0.85(6)\\
      & 5.0 &
      0.90(3) & 1.45( 5) & $-$0.73(1) & 0.83(3) & 0.78(3)\\
      & 3.0 &
      0.91(2) & 1.85( 4) & $-$0.82(1) & 0.82(2) & 0.76(2)\\
      & 2.6 &
      0.91(2) & 2.01( 4) & $-$0.86(1) & 0.81(2) & 0.74(2)\\
      & 2.1 &
      0.91(1) & 2.30( 5) & $-$0.93(1) & 0.79(1) & 0.71(2)\\
      \hline
      $\kappa$=$\kappa_s$ &
      $\infty$ &
      0.93(11) & 1.15(23) & $-$0.68(7) & 0.92(14) & 0.89(14)\\
      &10.0 &
      0.97( 9) & 1.22(17) & $-$0.69(5) & 0.91( 9) & 0.87( 9)\\
      & 5.0 &
      0.89( 4) & 1.43( 7) & $-$0.72(2) & 0.82( 4) & 0.77( 4)\\
      & 3.0 &
      0.90( 2) & 1.86( 6) & $-$0.82(2) & 0.80( 2) & 0.74( 2)\\
      & 2.6 &
      0.90( 2) & 2.03( 5) & $-$0.86(2) & 0.79( 2) & 0.73( 2)\\
      & 2.1 &
      0.90( 2) & 2.33( 6) & $-$0.93(2) & 0.78( 2) & 0.70( 2)\\
      \hline
      $\kappa$=$\kappa_c$ &
      $\infty$ &
      0.94(14) & 1.11(28) & $-$0.68(8) & 0.94(17) & 0.91(17)\\
      &10.0 &
      0.98(11) & 1.20(21) & $-$0.69(6) & 0.93(11) & 0.89(11)\\
      & 5.0 &
      0.88( 4) & 1.42( 9) & $-$0.72(2) & 0.81( 4) & 0.76( 5)\\
      & 3.0 &
      0.89( 3) & 1.87( 7) & $-$0.82(2) & 0.79( 3) & 0.73( 3)\\
      & 2.6 &
      0.89( 2) & 2.04( 6) & $-$0.85(2) & 0.78( 2) & 0.72( 3)\\
      & 2.1 &
      0.89( 2) & 2.35( 7) & $-$0.93(2) & 0.77( 2) & 0.69( 2)\\
    \end{tabular}
    \caption{Numerical results for $B_X^{lat}(a)$ and
             $B_P(m_b)$.} 
    \label{tab:B_X}
  \end{center}
\end{table}
\begin{table}
  \begin{center}
    \begin{tabular}{c|cc|cc}
     & \multicolumn{2}{c|}{One-loop}
     & \multicolumn{2}{c}{RG improved}\\
     & $q^*=\pi/a$ & $q^*=1/a$ & $q^*=\pi/a$ & $q^*=1/a$ \\ 
      \hline
     $B_B^{stat}(m_b)$
     & 0.90 & 0.88 & 0.88 & 0.88 \\
    \end{tabular}
    \caption{Numerical results for $B_B^{stat}(m_b)$. Statistical
    errors are omitted here.}  
    \label{tab:statB}
  \end{center}
\end{table}

\clearpage

\clearpage

\begin{thebibliography}{99}
\bibitem{CKM_unitarity}
  For a review of the constraints on the CKM matrix, see for example,
  C. Caso {\it et al.} (Particle Data Group),
  Eur. Phys. J. C \textbf{3}, 1 (1998).
\bibitem{LEPBOSC}
  LEP B Oscillations Working Group, LEPBOSC 98/3, available
  from http://www.cern.ch/LEPBOSC/ .
\bibitem{draper}
  T. Draper,
  plenary talk given at the XII International Symposium,
  Boulder, Colorado, to appear in Nucl. Phys. B,
  hep-lat/9810065.
\bibitem{NRQCD}
  B.A.~Thacker and G.P.~Lepage,
  Phys. Rev. D \textbf{43}, 196 (1991);
  G.P.~Lepage, L.~Magnea, C.~Nakhleh, U.~Magnea, and
  K.~Hornbostel, 
  Phys. Rev. D \textbf{46}, 4052 (1992).
\bibitem{Our}
  K-I.~Ishikawa, H.~Matsufuru, T.~Onogi, N.~Yamada, and
  S.~Hashimoto, Phys. Rev. D \textbf{56}, 7028 (1997).
\bibitem{EKM}
  A.X.~El-Khadra, A.S.~Kronfeld and P.B.~Mackenzie,
  Phys. Rev. D \textbf{55}, 3933 (1997).
\bibitem{sofarfb}
  S. Aoki {\it et al.} (JLQCD Collaboration),
  Phys. Rev. Lett. \textbf{80}, 5711 (1998);
  A. El-Khadra {\it et al.},
  Phys. Rev. D \textbf{58}, 014506 (1998);
  C. Bernard {\it et al.} (MILC Collaboration), 
  Phys. Rev. Lett. \textbf{81}, 4812 (1998).
\bibitem{jlqcd_nrfb}
  K-I. Ishikawa {\it et al.} (JLQCD Collaboration),
  talk given at the XII International Symposium, Boulder,
  Colorado, to appear in Nucl. Phys. B, hep-lat/9809152.
\bibitem{UK}
  A.K.~Ewing \textit{et al.} (UKQCD Collaboration),
  Phys. Rev. D \textbf{54}, 3526 (1996).
\bibitem{GM}
  V.~Gim\'enez and G.~Martinelli,
  Phys. Lett. B \textbf{398}, 135 (1997).
\bibitem{CDM}
  J.~Christensen, T.~Draper and C.~McNeile,
  Phys. Rev. D \textbf{56}, 6993 (1997).
\bibitem{BBM}
  C.~Bernard, T.~Blum and A.~Soni,
  Phys. Rev. D \textbf{58}, 014501 (1998) 
\bibitem{LL}
  L.~Lellouch and C.-J.D.~Lin (UKQCD Collaboration),
  talk given at the XII International Symposium,
  Boulder, Colorado, to appear in Nucl. Phys. B,
  hep-lat/9809018. 
\bibitem{FHH}
  J.M.~Flynn, O.F.~Hernandez and B.R.~Hill,
  Phys. Rev. D \textbf{43}, 3709 (1991).
\bibitem{BP}
  A.~Borrelli and C.~Pittori,
  Nucl. Phys. B \textbf{385}, 502 (1992); 
  an error in $D_R$ is corrected in \cite{PS,GR,IOY}.
\bibitem{PS}
  M.~Di~Pierro and C.T.~Sachrajda (UKQCD Collaboration),
  Nucl. Phys. B \textbf{534}, 373 (1998).
\bibitem{GR}
  V.~Gim\'enez and J.~Reyes,
  to appear in Nucl. Phys. B, hep-lat/9806023.
\bibitem{IOY}
  K-I.~Ishikawa, T.~Onogi and N.~Yamada,
  hep-lat/9812007.
\bibitem{YA}
  N. Yamada {\it et al.},
  talk given at the XII International Symposium, Boulder, Colorado,
  to appear in Nucl. Phys. B, hep-lat/9809156.
\bibitem{BBL}
  G.~Buchalla, A.J.~Buras and M.E.~Lautenbacher,
  Rev. Mod. Phys. \textbf{68}, 1125 (1996),
  and references therein.
\bibitem{SW}
  B.~Sheikholeslami and R.~Wholert,
  Nucl. Phys. B \textbf{259}, 572 (1985).
\bibitem{GH}
  M.~Golden and B.~Hill, 
  Phys. Lett. B \textbf{354}, 225 (1991).
\bibitem{Duncan}
  A.~Duncan {\it et al.},
  Phys. Rev. D \textbf{51}, 5101 (1995),
  and references therein.
\bibitem{LM}
  G.P.~Lepage and P.B.~Mackenzie,
  Phys. Rev. D \textbf{48}, 2250 (1993).
\bibitem{DT}
  C.T.H.~Davies and B.A.~Thacker,
  Phys. Rev. D \textbf{45}, 915 (1992);
\bibitem{MS}
  C.J.~Morningstar,
  Phys. Rev. D \textbf{48}, 2265 (1993).
\bibitem{Lepage91}
  G.P.~Lepage, 
  Nucl. Phys. (Proc. Suppl.) \textbf{26}, 45 (1992).
\bibitem{Hashimoto}
  S.~Hashimoto,
  Phys. Rev. D \textbf{50}, 4639 (1994).
\bibitem{Gimenez}
  V.~Gim\'enez, Nucl. Phys. B \textbf{401}, 116 (1993)
\bibitem{CFG}
  M.~Ciuchini, E.~Franco and V. Gim\'enez,
  Phys. Lett. B \textbf{388}, 167 (1996);
\bibitem{Buchalla}
  G.~Buchalla, Phys. Lett. B \textbf{395}, 364 (1997).
\end{thebibliography}
\end{document}